\documentclass[usegraphicx,usenatbib]{mn2e}
\usepackage{times}

\def\apj {ApJ}
\def\apjl {ApJL}
\def\apjs {ApJS}
\def\aj {AJ}
\def\aap {A\&A}
\def\mnras {MNRAS}
\def\nat {Nature}
\begin{document}
\title
[Evolution of galaxy colours in groups]
{The evolution of the bi-modal colour distribution of galaxies in SDSS groups}
\author[Mart\'{\i}nez, O'Mill \& Lambas]
{H\'ector J. Mart\'{\i}nez,$^{1,2}$\thanks{E-mail: julian@oac.uncor.edu}
Ana L. O'Mill$^1$ \& Diego G. Lambas$^{1,2}$\\
$^1$Grupo de Investigaciones en Astronom\'{\i}a Te\'orica y Experimental,
IATE, Observatorio Astron\'omico, Universidad Nacional de C\'ordoba,\\
Laprida 854, X5000BGR, C\'ordoba, Argentina \\
$^2$Consejo de Investigaciones Cient\'{\i}ficas y T\'ecnicas (CONICET),
Avenida Rivadavia 1917, C1033AAJ, Buenos Aires, Argentina\\}
\date{\today}
\pagerange{\pageref{firstpage}--\pageref{lastpage}} 
\maketitle
\label{firstpage}
\begin{abstract}
We analyse $u-r$ colour distributions for several samples of galaxies 
in groups drawn from the Fourth Data Release of the Sloan Digital Sky Survey. 
For all luminosity ranges and environments considered the colour 
distributions are well described by the sum of two Gaussian functions. 
We find that the fraction of galaxies in the red sequence is an increasing 
function of group virial mass.
We also study the evolution of the galaxy colour distributions at low redshift, 
$z\le0.18$ in the field and in groups for galaxies brighter than
$M_r-5\log(h)=-20$, finding significant evidence of recent 
evolution in the population of galaxies in groups. 
The fraction of red galaxies monotonically increases with decreasing 
redshift, this effect implies a much stronger evolution of galaxies in groups 
than in the field.
\end{abstract}
\begin{keywords}
galaxies: fundamental parameters -- galaxies: clusters: general --
galaxies: evolution 
\end{keywords}
\section{Introduction} 
The galaxy population in the local universe consists broadly of two classes of objects, 
early and late types, distinguishable by their morphology, broadband colour 
and star formation rate (e.g., \citealt{strat01,brinch04,balogh04b,kauffmann04}). 
The properties of early-type galaxies are almost independent of the environment
\citep{dress97,bernardi03} and there is evidence that the properties of 
late-types are insensitive to the environment too \citep{biviano90,zmm06}.
There is also conclusive evidence that this bi-modality exists at least out to
$z\sim1$, and that the fractions of early and late types
are different compared to $z=0$ \citep{bell04,tanaka05}. 
More recently, \citet{driver} and \citet{allen06} conclude that galaxy bi-modality 
reflects the two-component nature of galaxies (bulge-disc) rather than two 
distinct galaxy populations.

There are several physical processes related to environment that can be 
responsible of the observed bi-modality by transforming
galaxies from late to early types and by truncating their SFRs.
Some of them are typical of cluster environment, such as ram pressure 
\citep{gg72}, galaxy harassment \citep{moore96} and interactions with 
the cluster potential \citep{bv90}. Some other processes such as galaxy 
mergers and interactions should be more common in groups of galaxies where 
the relative velocities of the galaxies are lower. Another process that 
increases the fraction of red galaxies in groups or clusters is strangulation 
\citep{balogh00}, that consists in the removal of the hot gas reservoir of 
in-falling galaxies so their star formation halts after 
their cold gas is consumed.

In the last years, with the advent of large galaxy redshift surveys such
as the Sloan Digital Sky Survey (SDSS; \citealt{sdss}) and the Two-degree Field
Galaxy Redshift Survey \citep{2df}, several authors have studied the relation
between different galaxy properties and the environment
(e.g., \citealt{lewis02,mardom02,gomez03,goto04,tanaka04,balogh04a,balogh04b}).
In particular, \citet{balogh04b} found that, at fixed luminosity the fraction
of red galaxies is a strong function of projected galaxy density. 
\citet{baldry04} and \citet{balogh04b} have proposed a scenario where mergers are 
driving the bi-modality, with a red population resulting from merger processes,
and a blue population that form stars at a rate determined by their internal 
physical properties. In this scenario, to preserve the Gaussian nature of 
the colour distributions the environmentally triggered transformations 
from blue to red colours should occur in a short timescale, or at high redshift. 

Most studies have parametrised environment with the projected density
of galaxies brighter than a given luminosity threshold, measured typically using the 
area containing the 5-10th nearest neighbour. As pointed out by \citet{weinmann06},
the physical meaning of this projected density depends on the environment, 
while it measures local density in clusters, in low density regions a
more global density estimate is derived by this measurement. There are few studies 
that have investigated how galaxy properties correlate with halo mass using large 
group catalogues. \citet{yo02} found that the fraction of early type galaxies 
increases continuously from the lowest to the highest mass groups in the \citet{mz02} 
catalogue constructed from the 100K release of 2dFGRS. \citet{yang05b} confirmed 
this result using an independent group catalogue based in the final release of 2dFGRS.
However, \citet{tanaka04} found no dependence of SFR and morphology on group
velocity dispersion, $\sigma$, when analysing galaxies in groups identified 
in the first data release of SDSS. Consistently with this, \citet{balogh04b}
found no trend of the fraction of red galaxies with $\sigma$ in clusters.
Some of the most recent works on the subject agree that galaxy properties and halo 
mass are indeed correlated. \citet{weinmann06} analyse the dependence of colour, 
star formation and morphology on halo mass in a group catalogue constructed 
from the second data release of SDSS using the algorithm by \citet{yang05a}. 
By splitting galaxies into early, intermediate and 
late-types according to their colour and specific star formation rate, 
they find that at fixed luminosity, the fraction of early type galaxies is a 
smooth increasing function of halo mass. \citet{hm2} have shown that colour 
is the galaxy property that correlates best with group mass using the group 
catalogue by \citet{zmm06} constructed from the fourth data release of SDSS 
(DR4; \citealt{dr4}).

For a better understanding of the impact of group environment on galaxy evolution,
it is important to trace their redshift evolution. In clusters of galaxies, 
a strong evolution in the fraction of blue galaxies was originally detected by 
\citet{bo78} and later by other authors 
(e.g. \citealt{bo84,rakos95,margocav00,margoniner01,depropris03}).
Evolution of the fraction of galaxies of different morphological types has 
also been found in clusters
(e.g \citealt{dress97,andreon97,couch98,fasano00}).
Evolution in groups was reported by \citet{all-smith93}. They compared
a sample of groups photometrically selected in the vicinity of bright
radio galaxies at low ($z\leq0.25$) and intermediate ($0.25\leq z\leq0.5$) redshift
and report evolution of the blue galaxy fraction analogous to that observed in clusters.
However, field contamination is a significant limitation of photometric data,
and it is not clear how the radio selection might bias the sample
of groups. 
Robust evidence of the evolution of galaxies in groups was found
by \citet{wilman05} by comparing an intermediate redshift sample at $0.3\leq z\leq0.55$ 
from the CNOC2 survey \citep{carlberg01} with local groups ($0.05\leq z\leq0.1$) in 
the 2PIGG catalogue \citep{eke04}. The authors found that the fraction of passive
galaxies is a strong function of environment and luminosity and declines strongly
with redshift. Their results provide indications of the effect of
different mechanisms acting in high and intermediate density regions.

Making use of the large amount of galaxy data made publicly available by the 
Sloan Digital Sky Survey team in their Fourth Data Release,
we study in this paper how the bi-modality in the $u-r$ colour 
distribution of galaxies varies from field galaxies to group galaxies of 
different masses and seek for possible evolution in the nearby universe 
$z\leq 0.18$. This paper is organised as follows: in section 2 we describe 
the sample of  galaxies in groups we use; in section 3 we analyse 
the dependence on group virial mass of the $u-r$ colour distribution 
for several luminosity defined subsamples of galaxies, its evolution at 
low redshift is analysed in section 4. We summarise our results and discuss 
their implications in section 5.
\section{The samples}
The samples of galaxies used in this paper are included in the Main Galaxy 
Sample (MGS; \citealt{mgs}) of DR4. The sample of galaxies in groups was 
constructed by \citet{zmm06}. They identified groups of galaxies in the MGS 
of DR4 using the same technique as \citet{mz05}. The technique consists in a 
standard friend-of-friend algorithm for group identification together with 
a procedure to avoid the artificial merging of smaller systems in high density 
regions and an iterative method to compute reliable group centre positions.
The resulting group sample includes 14004 galaxy groups with at least 4 members
in the area spectroscopically surveyed by DR4, accounting for a total of 85728
galaxies. Our sample of field galaxies consists of those MGS DR4 galaxies that
were not identified as belonging to groups by \citet{zmm06}. Thus, our field
sample includes some galaxies that belong to small groups that were
undetected given the characteristics of the group finding procedure.

Galaxy magnitudes were corrected for Galactic extinction following 
\citet{sch98}, absolute magnitudes were computed assuming $\Omega_0=0.3$, 
$\Omega_{\Lambda}=0.7$ and $H_0=100~h~{\rm km~s^{-1}~Mpc^{-1}}$ 
and $K-$corrected using the method of \citet{blanton03}~({\small KCORRECT}
version 4.1). All magnitudes are in the AB system.
For the purpose of this work we use both Petrosian and Model magnitudes.
Since the MGS is defined using Petrosian magnitudes, we use them to define volume-limited 
subsamples of galaxies. For analysing $u-r$ colours of galaxies we use Model 
magnitudes instead, since aperture photometry may include non-negligible Poisson 
and background subtraction uncertainties in the $u-$band.

\begin{figure*}
\includegraphics[width=170mm]{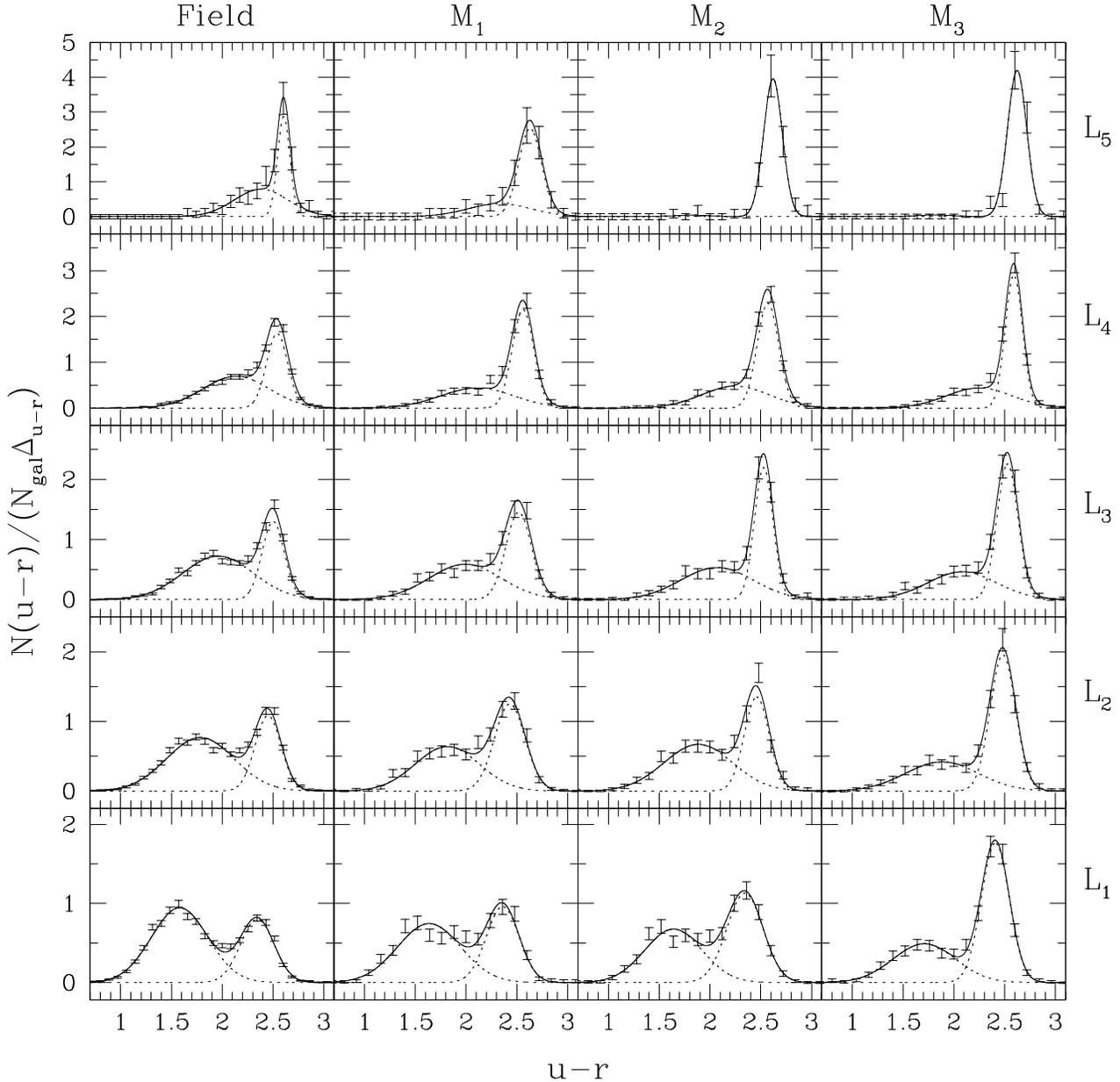}
\caption{
The $u-r$ model colour distributions for different subsamples of galaxies
at $z\leq0.055$. Each panel shows the colour distribution for the indicated 
environment (top axis) and luminosity (right axis).
Error-bars were computed assuming Poissonian statistics. We show in solid line
the best-fitting function that results from the sum of the two Gaussian functions
shown in dotted lines. 
}
\label{z0}
\end{figure*}
\section{Dependence of the colour distribution on group mass}
The colour distribution of galaxies at a given luminosity 
is well described by the sum of two Gaussian distributions 
(e.g., \citealt{balogh04b,baldry04}) representing the blue and the red sequences.
According to \citet{balogh04b}, at fixed luminosity, 
the mean colours of both populations are nearly independent of the galaxy 
surface density. In contrast, the fraction of galaxies in the red population 
strongly correlates with surface density but not with cluster velocity dispersion. 
This seems to be difficult to reconcile with the results by \citet{yo02}, 
\citet{yang05b} and \citet{weinmann06}.
In order to study in detail the evidence for differences in galaxy evolution in 
the group group environment and shed light on the processes governing this evolution,
in this section we explore the colour bi-modality of galaxies in groups of different 
virial masses and compare it to the corresponding colour distribution of field galaxies.

For studying the $u-r$ colour distribution of galaxies in groups
and in the field, we restrict our samples to $z\leq 0.055$. This choice guarantees
a volume limited sample of galaxies down to $M_r-5\log(h)=-19.0$. 
We then divide the galaxies into 5 luminosity bins: 
\begin{itemize}
\item $L_1:-19.5\leq M_r-5\log(h)\leq -19.0$;  
\item $L_2:-20.0\leq M_r-5\log(h)\leq -19.5$;  
\item $L_3:-20.5\leq M_r-5\log(h)\leq -20.0$;  
\item $L_4:-21.5\leq M_r-5\log(h)\leq -20.5$; \ and
\item $L_5:M_r-5\log(h)\leq -21.5$.  
\end{itemize}
In order to characterise the dependence of the distributions on group
mass, we divide each luminosity-defined subsample of galaxies in groups
into 3 mass bins according to the parent group virial mass:
\begin{itemize}
\item $M_1:11.0\leq \log(M h/M_{\odot})\leq 13.0$;  
\item $M_2:13.0\leq \log(M h/M_{\odot})\leq 13.5$ \ and  
\item $M_3:13.5\leq \log(M h/M_{\odot})\leq 15.0$ 
\end{itemize}
For each subsample, we have computed the $u-r$ colour distribution and 
fitted with the sum of two Gaussian functions: 
\begin{eqnarray}
\label{2gauss}
\frac{N(u-r)}{N_{\rm gal}\Delta_{u-r}}&=&
A_b\exp\left(-\frac{((u-r)-c_b)^2}{\sigma_b^2}\right)~ +\\
 & & A_r\exp\left(-\frac{((u-r)-c_r)^2}{\sigma_r^2}\right),\nonumber
\end{eqnarray}
where the sub-indexes $b$ and $r$ stand for `blue' and `red' sequences, 
$N_{\rm gal}$ is the number of galaxies in the subsample, and $\Delta_{u-r}$ 
is the colour bin's width. The fitting procedure consists in a standard 
Levenberg-Marquardt method that estimates the 6 best-fitting parameters 
in equation \ref{2gauss}.

The colour distribution for each subsample and its corresponding best-fitting
two-Gaussian function are shown in Figure \ref{z0}. 
We confirm previous findings that a two-Gaussian model is a good 
parametrisation of the observed colour distribution in all cases. 
Clearly, the colour distribution depends on both luminosity and environment. 
It is noticeable that the peak of the red galaxy population becomes more prominent 
in groups of increasing mass and for brighter galaxies. 
Another feature that varies among the different subsamples,
is the amplitude and position of the blue peak. It differs significantly between
field and groups. This peak weakens with mass and reddens from fainter to brighter 
galaxies and from low to high group mass.  

\begin{figure}
\includegraphics[width=90mm]{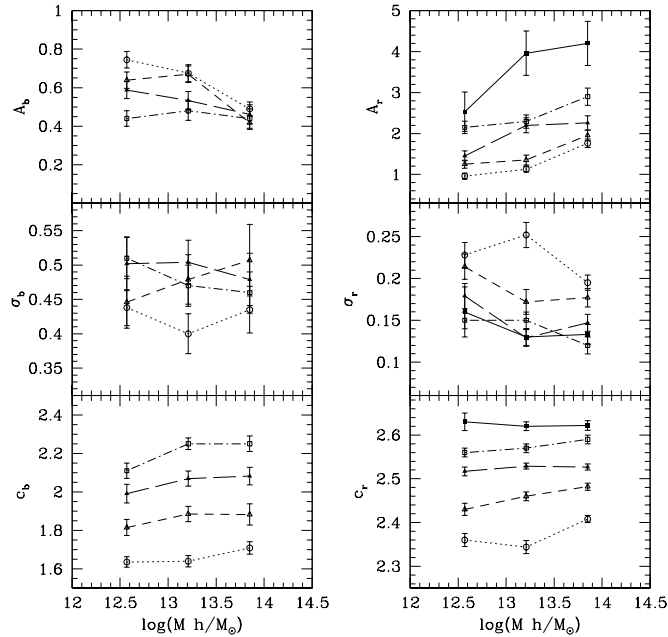}
\caption{
The double Gaussian best-fitting parameters of Figure \ref{z0} as a function of 
group mass for each luminosity defined subsample of galaxies. 
The $x$-axis values are the median of the group masses for each mass bin. 
Left column shows in its three panels the
values for the Gaussian function that represents the blue population,
while right column displays the corresponding values for the red population.
In both columns, top panels show the amplitudes $A_b$ and $A_r$, middle panels
the widths $\sigma_b$ and $\sigma_r$, and bottom panels 
the colours $c_b$ and $c_r$. 
Open circles and dotted lines correspond to $L_1$ galaxies, 
open triangles and short dashed lines to $L_2$ galaxies,
filled triangles and long dashed lines to $L_3$ galaxies,
open squares and dot-long dashed lines to $L_4$, galaxies and
filled squares and continuous lines to $L_5$ galaxies.
Notice the lack of blue population galaxies in the highest luminosity bin.
}
\label{z0mass}
\end{figure}

\begin{figure}
\includegraphics[width=90mm]{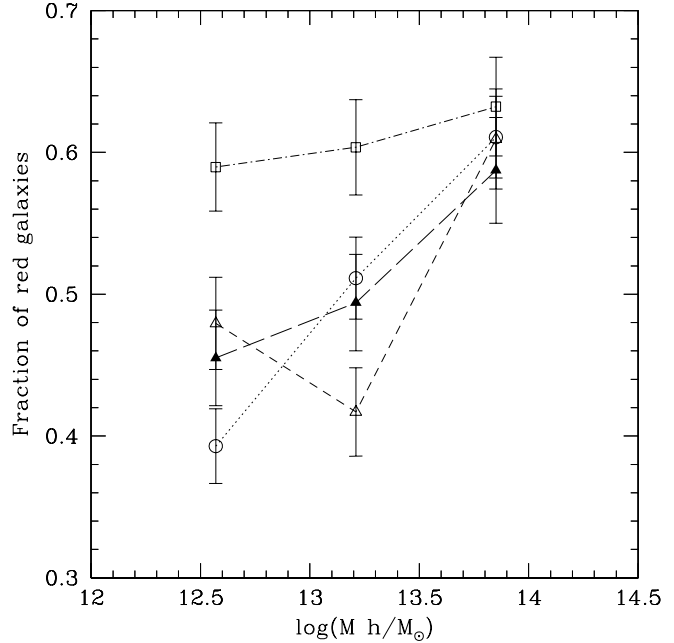}
\caption{
Fraction of galaxies in the red distribution as a function of group mass
Open circles and dotted lines correspond to $L_1$ galaxies, 
open triangles and short dashed lines to $L_2$ galaxies,
filled triangles and long dashed lines to $L_3$ galaxies, and
open squares and dot-long dashed lines to $L_4$, galaxies.
}
\label{RFM}
\end{figure}

In Figure \ref{z0mass} we show the best-fitting parameters corresponding to 
these distributions as a function of group virial mass. Notice that we do not show
the parameters corresponding to the blue sequence for $L_5$ galaxies since this sequence 
is absent in groups of masses $M_2$ and $M_3$, as can be seen in Figure \ref{z0}.
The parameters' trends become clearer in this figure. As a function of group mass
and for all luminosities considered here, we find that:
\begin{itemize}
\item the amplitude of the blue sequence, $A_b$, decreases, the width $\sigma_b$ is 
broadly consistent with no variation, and the mean colour, $c_b$, reddens;
\item on the other hand, the amplitude of the red sequence, $A_r$, strongly increases,
$\sigma_r$ decreases, and the mean colour $c_r$ reddens, except for $L_5$ galaxies
in which case it is consistent with no variation with mass.
\end{itemize}
In Figure \ref{RFM} we show the fraction of galaxies in the red sequence 
as a function of group virial mass, for $L_{1...4}$ galaxies. As pointed out
above, the colour distribution of $L_5$ galaxies is consistent with no
blue sequence for the higher mass bins.
Clearly, the fraction of red galaxies 
is a growing function of group mass for the remaining luminosities.
Over the whole range of masses considered here, the fraction of galaxies in the red 
sequence grows by $22\%$ for $L_1$, $13\%$ for $L_2$, $13\%$ for $L_3$, and
$4\%$ for $L_4$ galaxies.

Our findings are in qualitative agreement with the results by \citet{yo02}, 
\citet{weinmann06}, and \citet{hm2}. On the other hand, they disagree with the results
by \citet{balogh04b} and \citet{tanaka04}, although these authors consider velocity 
dispersion instead of mass. However, unlike \citet{weinmann06}, we do find a dependence 
of the mean colour of both sequences with mass. 
This could be due to the fact that we use $u-r$ colour, instead of $g-r$,
and $u$-band flux is a much better indicator of star formation than $g-$band flux.
Also, different ways of splitting galaxies into early and late types could make 
the difference between \citet{weinmann06} results and ours.
The observed dependence of the mean colour of each sequence on group mass agrees with 
\citet{balogh04b} results, who find similar changes with local density,
particularly for the blue sequence.

We have repeated our analysis by fixing the Gaussian widths given their lack of 
strong variation with mass. The results are essentially the same, indicating the 
robustness of our conclusions. 
\section{Evolution of the colour distribution}
In this section we explore the presence of evolution of galaxy colours at low redshift
and its dependence on environment.
We restrict our samples to those luminosities that allow the construction of 
volume-limited subsamples of galaxies with a number of objects large enough that allow 
splitting into at least 3 bins in redshift, that, in turn, contain enough objects for 
a good statistics. We restrict our analysis to the low redshift evolution of the colour
of galaxies brighter than $M_r-5\log(h)=-20$, that is, $L_3$, $L_4$ and $L_5$ galaxies.
For each of these subsamples, we have computed the colour distributions within 
different redshift limits: $z\le0.09$ for $L_3$ galaxies, $z\leq0.12$ for
$L_4$, and $z\le0.18$ for the $L_5$ subsample. 
That is, we investigate the evolution of $L_{3,4,5}$ galaxies in the last
$1.4,~1.9$ and $2.6h^{-1}$ Gyr, respectively.
In order to have robust statistics we only use two mass subsamples:
the low mass subsample comprising galaxies in groups with virial masses
$M_{\rm vir}\leq 10^{13.5}~ h^{-1}~M_{\odot}$, and a high mass subsample
including those galaxies in groups with $M_{\rm vir}>10^{13.5}~ h^{-1}~M_{\odot}$. 
These two mass bins correspond approximately to dividing the sample at the 
peak of the group virial mass distribution.

\begin{figure}
\includegraphics[width=90mm]{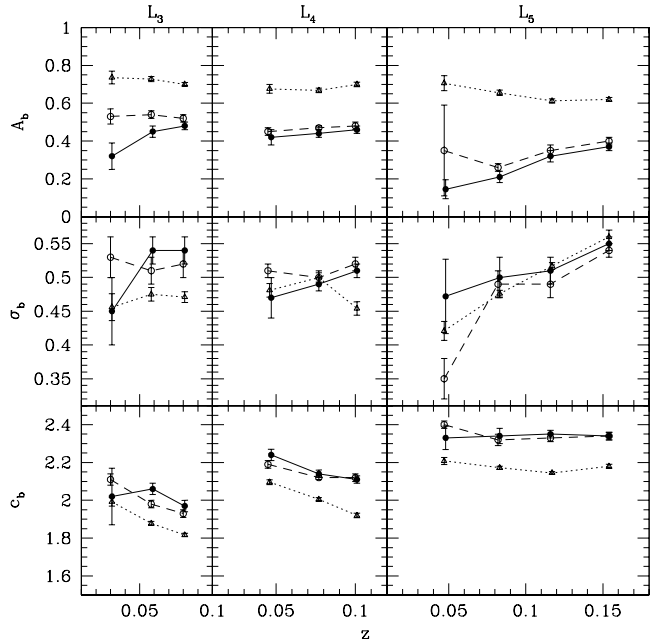}
\caption{
Best-fitting parameters for the blue population as a function of 
redshift for our luminosity samples $L_3$, $L_4$ and $L_5$ (see labels on top axis),
in the field (open triangles and dotted lines), all groups (open circles and dashed line)
and high mass groups (filled circles and continuous line).
The three top panels show the amplitude $A_b$, middle panels
the Gaussian width $\sigma_b$, and the bottom panels show the mean colour $c_b$.
}
\label{Bluez}
\end{figure}

\begin{figure}
\includegraphics[width=90mm]{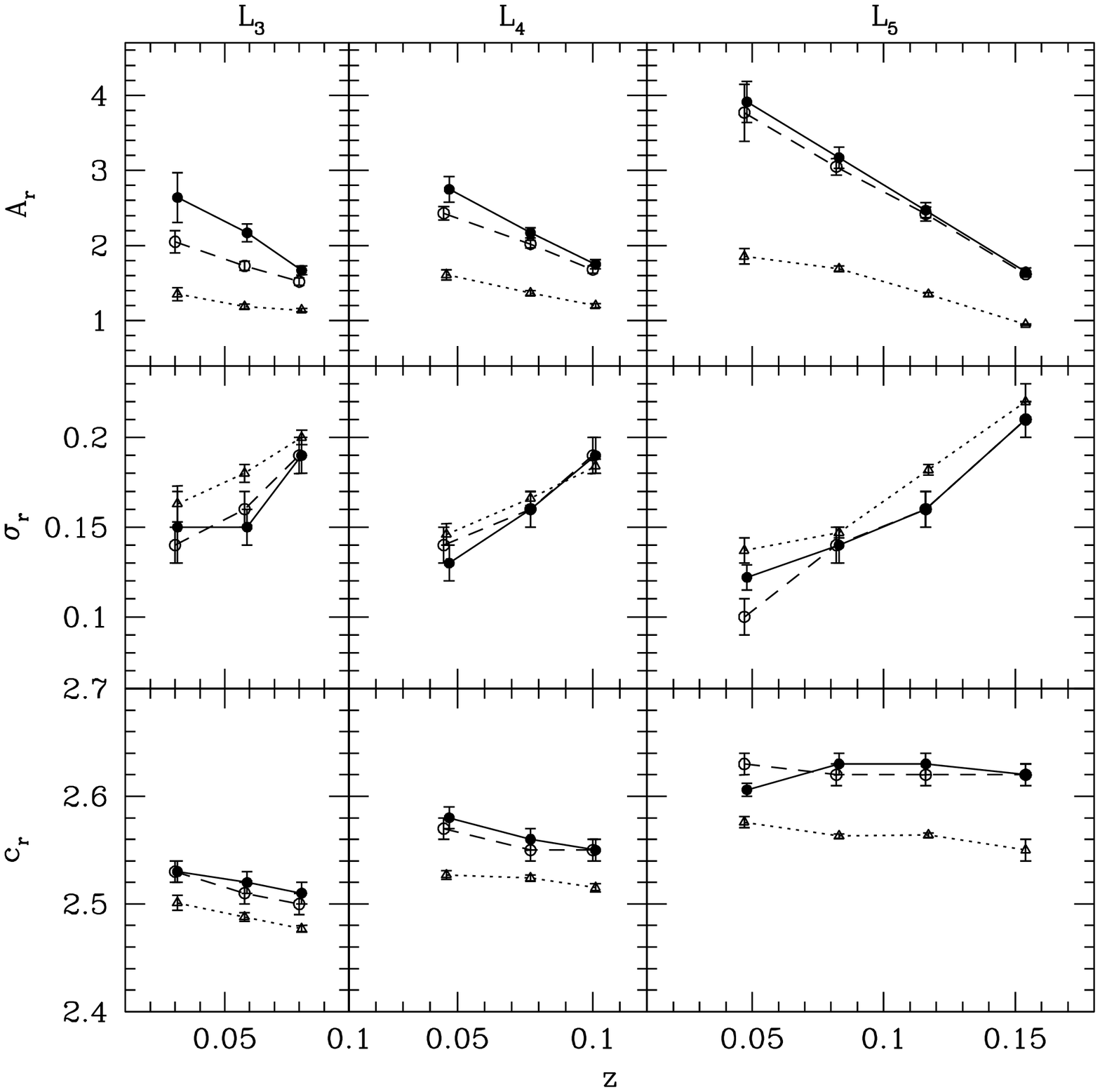}
\caption{
Best-fitting parameters for the red population as a function of 
redshift for our luminosity samples $L_3$, $L_4$ and $L_5$ (see labels on top axis),
in the field (open triangles and dotted lines), all groups (open circles and dashed line)
and high mass groups (filled circles and continuous line).
The three top panels show the amplitude $A_r$, middle panels
the Gaussian width $\sigma_r$, and the bottom panels show the mean colour $c_r$.
}
\label{Redz}
\end{figure}

We find in all cases that the two-Gaussian model is a good description of the 
colour distribution. We show the best-fitting parameters as a function of 
redshift for galaxies in the field, in groups and in the high mass subsample 
in Figures \ref{Bluez} and \ref{Redz}. Along with the trends found for luminosity 
and group mass in the previous subsection, there appear here some interesting 
trends as a function of redshift. 
Regarding the blue sequence (Figure \ref{Bluez}), the most significant change is that
in massive groups the amplitude increases with redshift.
The trends for $\sigma_b$ are noisy and in most cases consistent with no evolution
with the exception of $L_5$ galaxies, for which the sequence gets broader with 
increasing redshift.
The mean colour $c_b$ evolves with $z$ for $L_3$ and $L_4$ galaxies, being bluer
in the past.
The evolution with redshift is much stronger for the red sequence (Figure \ref{Redz}).
The amplitude of the red sequence is a decreasing function of redshift, much more
prominent in galaxy groups than in the field. In all cases the red sequence gets 
broader with redshift, and its mean colour gets slightly bluer.

The colour distribution of galaxies in groups significantly differs from that 
corresponding to field galaxies for all redshifts considered.
It should be taken into account that an important fraction of red galaxies in field 
samples are actually galaxies belonging to small groups that might have not 
been identified  by \citet{zmm06} because some of the other group members are fainter 
than the limiting apparent magnitude of the SDSS spectroscopic survey. 
Therefore, the actual differences between field
and group should be even more significant for the higher redshift bins.
We have tested the stability of our results by restricting the groups to a subsample
with at least 6 members. The analysis of this subsample gives
essentially the same results than that of the total group sample indicating
the lack of low number statistic biases as well as possible dependences on the number 
of members.

\begin{figure}
\includegraphics[width=90mm]{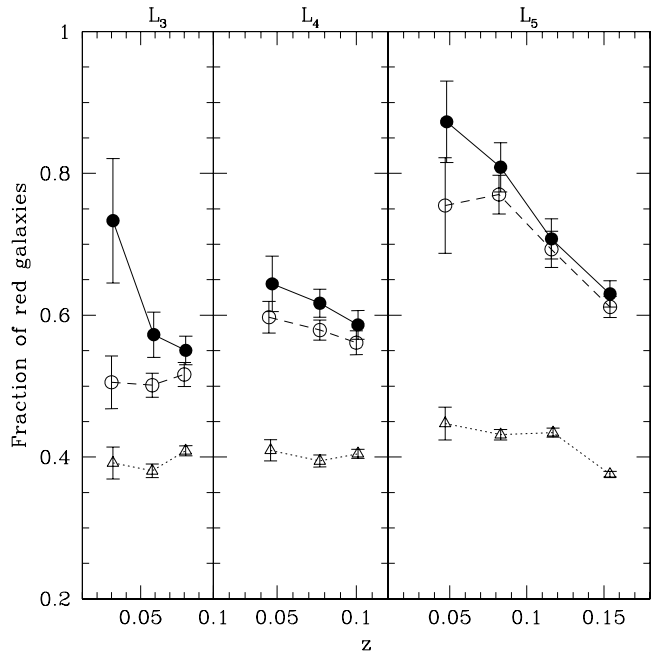}
\caption{
Fraction of galaxies in the red sequence as a function of redshift.
Open triangles and dotted lines: field galaxies; open circles and 
dashed lines: group galaxies; filled circles and continuous lines: galaxies in 
high mass groups.
}
\label{RF}
\end{figure}

In Figure \ref{RF} we show the fraction of galaxies in the red sequence as a function
of redshift. It is clear that in groups these fractions increase with cosmic time
even when we are considering a small redshift range. 
We also notice the remarkable different behaviour of field and massive group galaxies.
While the former show almost no changes, group galaxies exhibit a very significant 
decrease of the red population towards higher redshifts. This result is in agreement with
\citet{wilman05} comparison of nearby groups from the 2PIGG catalogue, at 
$z\sim 0.1$, with groups at $0.3\le z\le 0.55$ from the CNOC2 survey.
It is worth emphasising that in this work we have found statistically significant 
evolution in the recent ($<2.6h^{-1}$Gyr) past, made possible by the improved statistics
of the larger number of galaxies in DR4.
 
\section{Summary and conclusions}
We have used one of the largest sample of groups available, identified in the 
SDSS DR4 by \citet{zmm06} to study the colour distribution
of galaxies in groups, its dependence on luminosity and group mass, and its
evolution at low, $z\leq0.18$, redshift. 

We have found that, for all subsamples of galaxies in groups analysed, 
the $u-r$ model colour distribution is well fitted by
the sum of two Gaussian distributions that can be used to divide the galaxies
into a blue and a red population. The colour distributions strongly depend on 
galaxy luminosity and on parent group virial mass.
We have also found that the fraction of galaxies in the red sequence is a 
function of group mass
 
For galaxies brighter than $M_r-5\log(h)\leq-20$, we have studied the
evolution of the colour distribution at low, $z\leq0.18$, redshift, 
finding significant evidence of an increase on the fraction of red galaxies 
with cosmic time in groups in the last $\sim2.6h^{-1}$ Gyr. 
This effect is stronger for groups more massive than 
$M > 10^{13.5}h^{-1}M_{\odot}$, in stark contrast to the lack of evolution
observed in field galaxies over the same period. Our results are consistent
with the idea that the global evolution of galaxies (for example the observed decline
of the SFR since $z\sim1$; e.g., \citealt{cowie96})
takes place primarily in high density, dynamically evolved regions such 
as groups and clusters.

We have presented evidence that the processes that transform galaxies from late to 
early types have been more effective in groups of increasing mass than in the field, 
and that they have been efficiently acting on galaxies in the last $\sim2.6h^{-1}$ Gyr.
These findings could be used to constrain semi-analytic of galaxy formation,
providing important clues to the mechanisms driving the observed colour evolution.
As derived from our work the $u-r$ colour distribution is well
described by the sum of two Gaussian functions for the range of galaxy
luminosities and host group mass analysed within the redshift range
explored. To preserve this form, processes that transform galaxies should 
occur in short timescales, as discussed by 
\citet{balogh04b} (see also \citealt{baldry04}). In these studies, the authors propose
that the main process driving this evolution is merging. Groups are probably the best
environment for mergers given the high density and the relatively low galaxy velocity
dispersion. The results by \citet{zmm06},  
give additional support to this idea by showing that the characteristic luminosity 
of galaxies in groups is an increasing function of halo mass and that this 
behaviour is due to changes in the characteristic luminosity of galaxies in the red 
sequence. 

We conclude that there has been a significant difference in galaxy
colour evolution in groups in the last $\sim3$Gyr.
The observed evolution is stronger in the more massive galaxy groups, where
the relative fraction the most luminous galaxies
($M_r-5\log(h)\leq-21.5$) in the red sequence increases from $\sim 60\%$ to $90 \%$, 
compared to a roughly constant fraction of $40\%$ in the field.
\section*{Acknowledgements}
We thank Richard Bower for fruitful discussions and the anonymous referee
for helpful comments that improved this paper.
This work has been partially supported with grants from Consejo Nacional
de Investigaciones Cient\'\i ficas y T\'ecnicas de la Rep\'ublica Argentina
(CONICET), Secretar\'\i a de Ciencia y Tecnolog\'\i a de la Universidad 
de C\'ordoba, and Agencia Nacional de Promoci\'on Cient\'\i fica y Tecnol\'ogica, 
Argentina.

Funding for the Sloan Digital Sky Survey (SDSS) has been provided by the 
Alfred P. Sloan 
Foundation, the Participating Institutions, the National Aeronautics and Space 
Administration, the National Science Foundation, the U.S. Department of Energy, 
the Japanese Monbukagakusho, and the Max Planck Society. The SDSS Web site is 
http://www.sdss.org/.
The SDSS is managed by the Astrophysical Research Consortium (ARC) for the 
Participating Institutions. The Participating Institutions are The University 
of Chicago, Fermilab, the Institute for Advanced Study, the Japan Participation 
Group, The Johns Hopkins University, the Korean Scientist Group, Los Alamos 
National Laboratory, the Max Planck Institut f\"ur Astronomie (MPIA), the 
Max Planck Institut f\"ur Astrophysik (MPA), New Mexico State University, 
University of Pittsburgh, University of Portsmouth, Princeton University, 
the United States Naval Observatory, and the University of Washington.

\label{lastpage}
\end{document}